\documentclass[12pt]{article}

\usepackage{latexsym}
\usepackage{amssymb}
\usepackage{amsbsy}
\usepackage{amsmath}
\usepackage{epsfig}
\usepackage{graphics,color}
\usepackage{a4}
\usepackage{subfigure}
\usepackage{cite}

\newcommand{\bra}{\langle}
\newcommand{\ket}{\rangle}
\newcommand{\vecx}{{\mathbf x}}
  
\newcommand{\rmR}{{\rm R}}
\newcommand{\rmI}{{\rm I}}

\newcommand{\be}{\begin{equation}}
\newcommand{\ee}{\end{equation}}
\newcommand{\bea}{\begin{eqnarray}}
\newcommand{\eea}{\end{eqnarray}}
\newcommand{\bean}{\begin{eqnarray*}}
\newcommand{\eean}{\end{eqnarray*}}

\newcommand{\bit}{\begin{itemize}}
\newcommand{\eit}{\end{itemize}}
\newcommand{\half}{\frac{1}{2}}

\begin{document}

\title{ 
\vskip -100pt
{
\begin{normalsize}
\mbox{} \hfill \\
\mbox{} \hfill arXiv:1005.3468 [hep-lat]\\
\vskip  70pt
\end{normalsize}
}
\bf\large
On the convergence of complex Langevin dynamics:  the three-dimensional XY 
model at finite chemical potential}

\author{
\addtocounter{footnote}{2}
 Gert Aarts\thanks{email: g.aarts@swan.ac.uk} \;
 and 
 Frank A.\ James\thanks{email: pyfj@swan.ac.uk} \\
\addtocounter{footnote}{1}
\mbox{} \\
 {\em\normalsize Department of Physics, Swansea University} \\
 {\em\normalsize Swansea, United Kingdom} \\
}

\date{May 19, 2010}
\maketitle

\begin{abstract}
 The three-dimensional XY model is studied at finite chemical potential 
using complex Langevin dynamics. The validity of the approach is probed at 
small chemical potential using imaginary chemical potential and continuity 
arguments, and at larger chemical potential by comparison with the world 
line method. While complex Langevin works for larger $\beta$, we find that 
it fails for smaller $\beta$, in the region of the phase diagram 
corresponding to the disordered phase.
 Diagnostic tests are developed to identify symptoms correlated with 
incorrect convergence. We argue that the erroneous behaviour at smaller 
$\beta$ is not due to the sign problem, but rather resembles dynamics 
observed in complex Langevin simulations of simple models with complex 
noise.
 \end{abstract}
\newpage

\section{Introduction}
\label{sec:intro}
\setcounter{equation}{0}

Field theories with a complex action are difficult to treat 
nonperturbatively, because the weight $e^{-S} = |e^{-S}|e^{i\varphi}$ in 
the partition function is not real. Standard numerical approaches based on 
a probability interpretation and importance sampling will then typically 
break down, which is commonly referred to as the sign problem. This is a 
pressing problem for QCD at nonzero baryon chemical potential, where a 
nonperturbative determination of the phase diagram in the plane of 
temperature and chemical potential is still lacking 
\cite{deForcrand:2010ys}. Several methods have 
been developed to explore at least part of the phase diagram
\cite{Fodor:2001au,Fodor:2002km,Allton:2002zi,Gavai:2003mf,deForcrand:2002ci,
D'Elia:2002gd,D'Elia:2009qz,Kratochvila:2005mk,Alexandru:2005ix,Ejiri:2008xt,
Fodor:2007vv},
 but in general these can only be applied in a limited region. Recent 
years have also seen an intense study of the sign problem in QCD and 
related theories, which has led to new formulations
 \cite{Anagnostopoulos:2001yb,Ambjorn:2002pz}
and considerable insight into how the 
complexity of the weight interplays with physical observables
\cite{Akemann:2004dr,Splittorff:2005wc,Splittorff:2006fu,Han:2008xj,
Bloch:2008cf,Lombardo:2009aw,Danzer:2009dk,Andersen:2009zm,
Hands:2010zp,Bringoltz:2010iy}.
 Finally, in some theories the sign problem can be eliminated completely, 
using a reformulation which yields a manifestly real and positive weight
 \cite{Chandrasekharan:1999cm,Endres:2006xu,Chandrasekharan:2008gp,
Banerjee:2010kc}.
 This demonstrates that the sign problem is not a problem of principle for 
a theory, but instead tied to the formulation and/or algorithm. For QCD an 
exact reformulation without a sign problem has unfortunately not (yet) 
been found.

Complex Langevin dynamics \cite{Parisi:1984cs,Klauder:1983} 
offers the possibility of a general solution to this problem. In this 
formulation the fields, denoted here collectively as $\phi$, are 
supplemented with an additional fictional Langevin time $\vartheta$ and 
the system evolves according to the stochastic equation,
 \be
 \label{eqphi}
 \frac{\partial \phi_{x}(\vartheta)}{\partial \vartheta} = 
 -\frac{\delta S[\phi;\vartheta]}{\delta \phi_{x}(\vartheta)} 
 + \eta_x(\vartheta).
\ee
 In the case that the action is complex, the fields are 
\emph{complexified} as
\be
\phi\to\phi^\rmR + i\phi^\rmI,
\ee
 and the Langevin equations read (using general complex noise)
 \begin{subequations}
 \begin{align}
 \frac{\partial \phi_{x}^\rmR}{\partial \vartheta} &= K_{x}^\rmR + 
 \sqrt{N_\rmR}\eta^\rmR_{x},
 & K_{x}^\rmR = -\mbox{Re}\frac{\delta S}{\delta 
 \phi_{x}}\Big|_{\phi\to\phi^\rmR+i\phi^\rmI}, 
\\
 \frac{\partial \phi_{x}^\rmI}{\partial \vartheta} &= K_{x}^\rmI + 
\sqrt{N_\rmI}\eta^\rmI_x, 
& K_{x}^\rmI  = -\mbox{Im}\frac{\delta S}{\delta \phi_{x}}\Big|_{\phi\to\phi^\rmR+i\phi^\rmI}.
\end{align}
\label{eqphic}
 \end{subequations}
 The strength of the noise in the real and imaginary components of the 
Langevin equation is constrained via $N_\rmR-N_\rmI=1$, and the 
noise furthermore satisfies
 \begin{subequations}
\begin{align}
&\langle \eta_x^\rmR(\vartheta)\rangle = \langle \eta_x^\rmI(\vartheta)\rangle = \langle \eta^\rmR_{x}(\vartheta)\eta^\rmI_{y}(\vartheta') \rangle = 0,  \\
&\langle \eta^\rmR_{x}(\vartheta)\eta^\rmR_{y}(\vartheta') \rangle =
\langle \eta^\rmI_{x}(\vartheta)\eta^\rmI_{y}(\vartheta') \rangle 
= 2\delta_{xy}\delta(\vartheta - \vartheta'),
\end{align}
\end{subequations}
i.e., it is Gaussian.
 Since the complex action is only used to generate the drift terms but not 
for importance sampling, complex Langevin dynamics can potentially avoid 
the sign problem.\footnote{Early studies of complex Langevin 
dynamics can be found in, e.g., Refs.\ 
\cite{Klauder:1985b,Karsch:1985cb,Ambjorn:1985iw,Ambjorn:1986fz,
Flower:1986hv,Ilgenfritz:1986cd}. 
Ref.\ \cite{Damgaard:1987rr} contains a further guide to the 
literature. More recent work includes Refs.\ 
\cite{Berges:2005yt,Berges:2006xc,Berges:2007nr,Aarts:2008rr,
Aarts:2008wh,Aarts:2009hn,Aarts:2009dg,Aarts:2009uq,Pehlevan:2007eq,
Guralnik:2009pk}.
}

In the limit of infinite Langevin time, noise averages of observables
should 
equal the standard quantum expectation values.  For a real action/Langevin 
dynamics, formal proofs that observables converge to the correct value can 
be formulated, using properties of the associated Fokker-Planck equation
\cite{Damgaard:1987rr}. 
If the action is complex and the Langevin dynamics extends into the 
expanded complexified space, these proofs no longer hold. 
Nevertheless, a formal derivation of the validity of the approach can 
still be given, employing holomorphicity and the Cauchy-Riemann equations. 
We sketch here the basic notion, suppressing all indices for notational 
simplicity, and refer to Ref.\ \cite{Aarts:2009uq} for details.

 Associated with the Langevin process (\ref{eqphic}) is a (real and 
positive) probability density $P[\phi^\rmR, \phi^\rmI;\vartheta]$, which 
evolves according the Fokker-Planck equation
 \be
\frac{\partial P[\phi^\rmR,\phi^\rmI;\vartheta]}{\partial\vartheta} = L^T 
P[\phi^\rmR, \phi^\rmI;\vartheta],
\ee
with the Fokker-Planck operator
\be
L^T = \frac{\partial}{\partial\phi^\rmR} \left[ N_\rmR \frac{\partial}{\partial\phi^\rmR}- K^\rmR\right]
 +       \frac{\partial}{\partial\phi^\rmI} \left[ N_\rmI \frac{\partial}{\partial\phi^\rmI}- K^\rmI\right].
\ee
 Stationary solutions of this Fokker-Planck equation are only known in 
very special cases 
\cite{Aarts:2009hn,Ambjorn:1985iw,Aarts:2009uq,Nakazato:1985zj}. 
Expectation values obtained by solving the stochastic process should then 
equal
 \be
\label{eqP}
\bra O\ket_{P(\vartheta)} 
= \frac{\int D\phi^\rmR D\phi^{\rmI}\, P[\phi^\rmR,\phi^\rmI;\vartheta] 
O[\phi^\rmR+i\phi^\rmI]}
{\int D\phi^\rmR D\phi^\rmI\, P[\phi^\rmR,\phi^\rmI;\vartheta]}.
\ee
 However, we may also consider expectation values with respect to a 
complex weight $\rho[\phi;\vartheta]$,
\be
\bra O\ket_{\rho(\vartheta)} 
= \frac{\int D\phi\, \rho[\phi;\vartheta] O[\phi]}{\int D\phi\, 
\rho[\phi;\vartheta]},
\ee
 where, using Eq.\ (\ref{eqphi}),  $\rho$ evolves according to a complex 
Fokker-Planck equation
\be
\frac{\partial\rho[\phi;\vartheta]}{\partial\vartheta} = 
L_0^T\rho[\phi;\vartheta],
\;\;\;\;\;\;\;\;
L_0^T = \frac{\partial}{\partial\phi} \left[  \frac{\partial}{\partial\phi}+ \frac{\partial S}{\partial\phi} \right].
\ee
 This equation has the desired stationary solution $\rho[\phi]\sim 
\exp(-S)$.

 Under some assumptions and relying on holomorphicity and partial 
integration \cite{Aarts:2009uq}, one can show that these expectation 
values are equal, and
 \be
 \bra O\ket_{P(\vartheta)} =  \bra O\ket_{\rho(\vartheta)}.
 \ee 
If it can subsequently be shown that 
\be
 \lim_{\vartheta\to\infty} \bra O\ket_{\rho(\vartheta)} = \bra 
O\ket_{\rho(\infty)},
\;\;\;\;\;\;\;\; 
\rho(\phi;\infty)\sim \exp(-S),
\ee
 the applicability of complex Langevin dynamics is demonstrated. In Ref.\ 
\cite{Aarts:2009uq} this proposal was studied in some detail in the case 
of simple models. Remarkably it was found that for complex noise 
($N_\rmI>0$), the Langevin dynamics does {\em not} converge to the correct 
answer. On the other hand, for real noise ($N_\rmI=0$) correct convergence 
was observed.

In this paper, we continue our investigation into the applicability of 
complex Langevin dynamics at finite chemical potential \cite{Aarts:2008rr, 
Aarts:2008wh,Aarts:2009hn,Aarts:2009dg,Aarts:2009uq}. We consider the 
three-dimensional XY model for a number of reasons. We found earlier that 
this theory is very sensitive to instabilities and runaways and therefore 
requires the use of an adaptive stepsize \cite{Aarts:2009dg}. This is 
similar to the case of QCD in the heavy dense limit 
\cite{Aarts:2008rr,Aarts:2009dg}. As QCD, this theory has a Roberge-Weiss 
periodicity at imaginary chemical potential \cite{Aarts:2009dg,Roberge:1986mm}. 
Furthermore, it is closely related to the relativistic Bose gas at finite 
chemical potential, for which complex Langevin dynamics was shown to work 
very well (at weak coupling in four dimensions) 
\cite{Aarts:2008wh,Aarts:2009hn}. Finally, this theory can be rewritten 
using a world line formulation without a sign problem 
\cite{Chandrasekharan:2008gp,Banerjee:2010kc}, which can be solved 
efficiently using the worm algorithm \cite{Banerjee:2010kc,worm}. This 
allows for a direct comparison for all parameter values.

The paper is organized as follows. In Sec.\ \ref{sec:xymodel}, we remind 
the reader of some details of the XY model at real and imaginary chemical 
potential, the adaptive stepsize algorithm we use and the related 
phase-quenched XY model. The world line formulation and some properties of 
the strong-coupling expansion are briefly mentioned in Sec.\ \ref{sec:wl}. 
We then test the validity of complex Langevin dynamics in Sec.\ 
\ref{sec:comparison} and develop diagnostic tests in Sec.\ 
\ref{sec:diagnostics}. In the Conclusion we summarize our findings and 
discuss possible directions for the future.

\section{XY model}
\label{sec:xymodel}
\setcounter{equation}{0}

The action of the XY model at finite chemical potential is
\begin{equation}
 S = -\beta\sum_x\sum_{\nu=0}^2\cos(\phi_{x} - \phi_{x+\hat{\nu}} - 
i\mu\delta_{\nu,0}) ,
\label{eq:action}
\end{equation}
 where $0\leq \phi_x<2\pi$. The theory is defined on a lattice of size 
$\Omega=N_\tau N_s^2$, and we use periodic boundary conditions. The 
chemical potential $\mu$ is coupled to the Noether charge associated with 
the global symmetry $\phi_x\rightarrow\phi_x+\alpha$ and is introduced in 
the standard way \cite{Hasenfratz:1983ba}. The action satisfies 
$S^*(\mu)=S(-\mu^*)$. At vanishing chemical potential the theory is known 
to 
undergo a phase transition at $\beta_{c}=0.45421$ 
\cite{Campostrini:2000iw,Banerjee:2010kc} between a disordered phase when 
$\beta<\beta_{c}$ and an ordered phase when $\beta>\beta_{c}$.

The drift terms appearing in the complex Langevin equations are given by
 \begin{subequations}
\begin{align}
K^\rmR_x = -\beta\sum_{\nu} & \Big[ \sin(\phi_x^\rmR - \phi_{x+\hat{\nu}}^\rmR)\cosh(\phi_{x}^\rmI - 
\phi_{x+\hat{\nu}}^\rmI - \mu\delta_{\nu,0}) \nonumber\\
 &{}+ \sin(\phi_{x}^\rmR - \phi_{x-\hat{\nu}}^\rmR)\cosh(\phi_{x}^\rmI - \phi_{x-\hat{\nu}}^\rmI + \mu\delta_{\nu,0})  \Big], \\
K^\rmI_x = -\beta\sum_{\nu} & \Big[ \cos(\phi_{x}^\rmR - \phi_{x+\hat{\nu}}^\rmR)\sinh(\phi_{x}^\rmI - 
\phi_{x+\hat{\nu}}^\rmI - \mu\delta_{\nu,0}) \nonumber\\
 &{}+ \cos(\phi_{x}^\rmR - \phi_{x-\hat{\nu}}^\rmR)\sinh(\phi_{x}^\rmI - \phi_{x-\hat{\nu}}^\rmI + \mu\delta_{\nu,0})  \Big]. 
\end{align}
\end{subequations}
 The equations are integrated numerically by discretizing Langevin time as 
$\vartheta=n\epsilon_n$ with $\epsilon_n$ the adaptive stepsize. Explicitly, 
 \begin{subequations}
\begin{align}
\phi_{x}^\rmR(n+1) {}&= \phi_{x}^\rmR(n) + \epsilon_{n}K_{x}^\rmR(n) + \sqrt{\epsilon_{n}}\eta_{x}(n),\\
\phi_{x}^\rmI(n+1) {}&= \phi_{x}^\rmI(n) + \epsilon_{n}K_{x}^\rmI(n),
\end{align} 
\end{subequations}
 where we specialized to real noise, with 
$\bra\eta_x(n)\eta_{x'}(n')\ket=2\delta_{xx'}\delta_{nn'}$. In the case 
that 
$\mu=\phi^\rmI=0$, the drift terms are bounded and $|K_x^\rmR|<6\beta$. 
When $\phi^\rmI\neq 0$, the drift terms are unbounded, which can result in 
instabilities and runaways. In this particular theory, much care is 
required to numerically integrate the dynamics in a stable manner and we 
found that an adaptive stepsize is mandatory \cite{Aarts:2009dg}. At each 
timestep, the stepsize is determined according to
 \be
  \epsilon_{n} = \min\left\{\bar{\epsilon}, \bar{\epsilon}\frac{\langle 
K^{\rm max} \rangle}{K^{\rm max}_n}\right\} ,
 \ee
where
\be
 K^{\rm max}_n =  \max_x \left| K^\rmR_x(n) + i K^\rmI_x(n)\right|.
 \ee
 Here $\bar{\epsilon}$ is the desired target stepsize and $\langle K^{\rm 
max} \rangle$ is either precomputed or computed during the thermalisation 
phase. All observables are analyzed over equal periods of Langevin time to 
ensure correct statistical significance.

 The observable we focus on primarily in this study is the action density 
$\langle S\rangle/\Omega$. After complexification the action is written as 
$S=S^\rmR+iS^\rmI$, with
\begin{subequations}
\begin{align}
S^{\rmR} & = - \beta\sum_{x,\nu}\cos(\phi_{x}^\rmR - \phi_{x+\hat{\nu}}^\rmR)\cosh(\phi_{x}^\rmI - \phi_{x+\hat{\nu}}^\rmI - \mu\delta_{\nu,0}), \\
S^{\rmI} & =  \beta\sum_{x,\nu}\sin(\phi_{x}^\rmR - \phi_{x+\hat{\nu}}^\rmR)\sinh(\phi_{x}^\rmI - \phi_{x+\hat{\nu}}^\rmI - \mu\delta_{\nu,0}).
\end{align}
\end{subequations}
 After noise averaging, the expectation value of the imaginary part is 
consistent with zero while the expectation value of the real part is even 
in $\mu$, as is expected from symmetry considerations.


By choosing an imaginary chemical potential $\mu=i\mu_\rmI$ the action 
(\ref{eq:action})
becomes purely real. This has both the advantage of enabling standard 
Monte Carlo algorithms to be applied (we choose to employ real Langevin 
dynamics) and that the behaviour at $\mu^2\gtrsim0$ can be assessed by 
continuation from $\mu^2\lesssim0$. The action and drift term with 
imaginary chemical potential are
 \begin{align}
S_{\rm imag} =& -\beta\sum_{x,\nu}\cos(\phi_{x} - \phi_{x+\hat{\nu}} + \mu_\rmI\delta_{\nu,0}), \\
K_{x} =& -\beta\sum_{\nu}\left[ 
\sin(\phi_{x} - \phi_{x+\hat{\nu}} + \mu_\rmI\delta_{\nu,0}) + \sin(\phi_{x} - \phi_{x-\hat{\nu}} - \mu_\rmI\delta_{\nu,0}) \right] .
\end{align}
 This theory is periodic under $\mu_\rmI\to\mu_\rmI+2\pi/N_{\tau}$, which 
yields a Roberge-Weiss transition at $\mu_\rmI=\pi/N_{\tau}$, similar as in 
QCD \cite{Roberge:1986mm}.
 This periodicity can be made explicit by shifting the chemical potential 
to the final time slice, via the field redefinition 
$\phi_{\vecx,\tau}\longrightarrow\phi_{\vecx,\tau}^{\prime} = 
\phi_{\vecx,\tau}-\mu_\rmI\tau$. The action is then (for arbitrary complex 
chemical potential)
 \be
 S_{\rm fts} = -\beta\sum_{x,\nu}\cos(\phi_{x} - \phi_{x+\hat{\nu}} - 
 iN_{\tau}\mu\delta_{\tau,N_{\tau}}\delta_{\nu,0}).
\ee
 We have also carried out simulations with this action and confirmed the 
results obtained with the original formulation. The sole exception was the 
largest $\beta$ value ($\beta=0.7$), where the original action missed the 
Roberge-Weiss transition, while the final-time-slice formulation located 
it without problems.

The severity of the sign problem is conventionally 
(see e.g.\ Ref.\ \cite{deForcrand:2010ys})
estimated by the 
expectation value of the phase factor $e^{i\varphi}=e^{-S}/|e^{-S}|$ in 
the phase quenched theory, i.e.\ in the theory where only the real 
part of the action (\ref{eq:action}) is included in the Boltzmann weight.
 In this case, the phase quenched theory is the anisotropic XY model, with 
the action
\be
 S_{\rm pq} = -\sum_{x,\nu} \beta_\nu\cos(\phi_x-\phi_{x+\hat\nu}),
\ee
where $\beta_0=\beta\cosh\mu$, and $\beta_{1,2}=\beta$.


\section{World line formulation}
\label{sec:wl}
\setcounter{equation}{0}

The advantage of the XY model is that it can be formulated without a sign 
problem by an exact rewriting of the partition function in terms of world 
lines \cite{Chandrasekharan:2008gp,Banerjee:2010kc}.\footnote{The world 
line formulation has of course a long history in lattice gauge theory, see 
e.g.\ Ref.\ \cite{Banks:1977cc}. Recent work includes Refs.\ 
\cite{Wenger:2008tq,deForcrand:2009dh,Wolff:2009kp}. For a review, see Ref.\ 
\cite{Chandrasekharan:2008gp}.
 }
 Moreover, this dual formulation can be simulated efficiently with a worm 
algorithm \cite{worm,Banerjee:2010kc}, which allows us to compare the 
results obtained with complex Langevin dynamics with those from the world 
line approach. We briefly repeat some essential elements of the world line 
formulation and refer to Ref.\ \cite{Banerjee:2010kc} for more details. 
The partition function can be rewritten using the identity
 \be
 e^{\beta\cos\phi} = \sum_{k=-\infty}^{\infty}I_{k}(\beta)e^{ik\phi}, 
 \ee
 where $I_{k}(\beta)$ are the modified Bessel functions of the first kind. 
Using this replacement and integrating over the fields, the partition 
function is written as
 \begin{equation}
Z = \int D\phi \, e^{-S} = 
\sum_{[k]}\prod_{x,\nu} I_{k_{x,\nu}}(\beta)e^{k_{x,\nu}\mu\delta_{\nu,0}}
\delta\left(\sum_{\nu}\left[ k_{x,\nu} - k_{x-\hat{\nu},\nu} \right] \right) .
\end{equation}
 The sum over $[k]$ indicates a sum over all possible world line 
configurations. Since $\langle S\rangle = -\beta\frac{\partial \ln 
Z}{\partial \beta}$, the action can be computed from
 \begin{equation}
 \langle S \rangle = -\beta\left\langle 
 \sum_{x,\nu}\left[ 
 \frac{I_{k_{x,\nu}-1}(\beta)}{I_{k_{x,\nu}}(\beta)} - 
 \frac{k_{x,\nu}}{\beta}
 \right]\right\rangle_{\rm wl},
\end{equation}
 where the brackets denote the average over world line configurations. To 
compute this average, we have implemented the worm algorithm, following 
Ref. \cite{Banerjee:2010kc}. We note here, amusingly, that the world line 
formulation has a sign problem at imaginary chemical potential.

Inspired by Ref.\ \cite{Langelage:2010yn},
we have also studied a (low-order) strong-coupling expansion of this 
model, using
\be
I_k(2x) = \frac{x^k}{k!}\left(
1 + \frac{x^2}{k+1}
  + \frac{x^4}{2(k+2)(k+1)} +\ldots\right).
\ee
 At strong coupling the chemical potential cancels in most world lines, 
except when the world line wraps around the temporal direction. At leading 
order in the strong-coupling expansion, it then appears in the combination 
$(\half\beta e^\mu)^{N_\tau}$. In the thermodynamic limit it therefore contributes only 
when $\half\beta e^\mu \geq 1$. Hence a simple strong-coupling estimate 
for the critical coupling at nonzero $\mu$ is given by
 \be
 \beta_c(\mu) = 2e^{-\mu}.
 \ee
The $\mu$-independence at small $\beta$ and $\mu$ is known as the Silver 
Blaze feature in QCD \cite{Cohen:2003kd}.

 The partition function is expressed in terms of the free energy density 
$f$ as $Z=\exp(-\Omega f)$. A strong-coupling expansion to order $\beta^4$ 
on a lattice with $N_{\tau}>4$ yields
\be 
f = -\frac{3}{4}\beta^2-\frac{21}{64}\beta^4 + {\cal 
O}(\beta^6), 
\ee 
and hence 
\be 
\label{eqSsc}
\bra S\ket/\Omega = 
-\frac{3}{2}\beta^2 - \frac{21}{16}\beta^4 + {\cal O}(\beta^6).
\ee 
In the phase quenched theory we find
\be
f_{\rm pq} = 
-\frac{1}{4}\beta^2\left(2+\cosh^2\mu\right)
-\frac{1}{64}\beta^4\left(14+8\cosh^2\mu-\cosh^4\mu\right) 
+ {\cal O}(\beta^6).
\ee
We can now estimate the severeness of the sign problem at strong coupling. 
The average phase factor takes the standard form,
\be
\bra e^{i\varphi}\ket_{\rm pq} = \frac{Z}{Z_{\rm pq}} = 
\exp\left[-\Omega\Delta f\right],
\;\;\;\;\;\;\;\;
\Delta f = f-f_{\rm pq},
\ee
where in this case
\be
\Delta f = 
\frac{1}{4}\beta^2\left(\cosh^2\mu-1\right)
+\frac{1}{64}\beta^4\left(\cosh^2\mu-1\right)\left(7-\cosh^2\mu\right) 
+ {\cal O}(\beta^6).
\ee
 On a finite lattice and for small chemical potential we find therefore 
the sign problem to be mild in the strong-coupling limit, since the volume 
factor is balanced by $\beta^2\mu^2/4\ll 1$.


\section{Comparison}
\label{sec:comparison}
\setcounter{equation}{0}

\begin{figure}[htp]
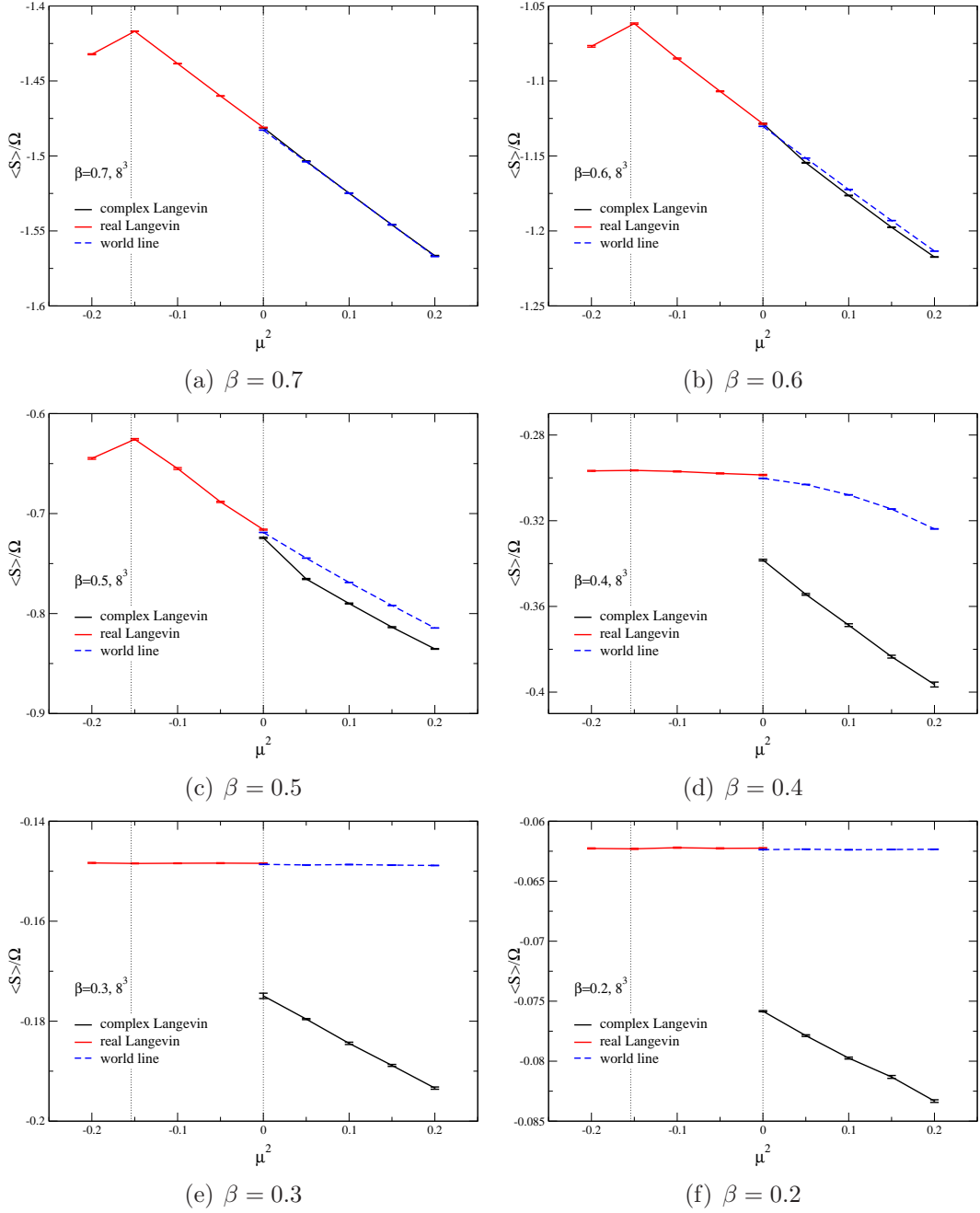

\centering
\subfigure[$\beta=0.7$]{
\includegraphics[width=0.47\textwidth]{act_b0.7_x8.eps}
}
\subfigure[$\beta=0.6$]{
\includegraphics[width=0.47\textwidth]{act_b0.6_x8.eps}
}
\subfigure[$\beta=0.5$]{
\includegraphics[width=0.47\textwidth]{act_b0.5_x8.eps}
}
\subfigure[$\beta=0.4$]{
\includegraphics[width=0.47\textwidth]{act_b0.4_x8.eps}
}
\subfigure[$\beta=0.3$]{
\includegraphics[width=0.47\textwidth]{act_b0.3_x8.eps}
}
\subfigure[$\beta=0.2$]{
\includegraphics[width=0.47\textwidth]{act_b0.2_x8.eps}
}
\caption{
 Real part of action density $\bra S\ket/\Omega$ as a function of
$\mu^2$ on a lattice of size $8^3$, using complex Langevin dynamics and
the world line formulation at real $\mu$ ($\mu^2>0$) and real Langevin
dynamics at imaginary $\mu$ ($\mu^2<0$). The vertical lines on the left
indicate the Roberge-Weiss transitions at $\mu_\rmI=\pi/8$.
 }
\label{fig:action-plots}
\end{figure}

We start to assess the applicability of complex Langevin dynamics for this 
model at small chemical potential. In this case we can use continuity 
arguments to compare observables at real and imaginary chemical potential. 
In Fig.\ \ref{fig:action-plots} the real part of the action density is 
shown as a function of $\mu^2$, for several values of $\beta$: from the 
ordered phase at large $\beta$ to the disordered phase at low $\beta$. We 
observe that at the highest values of $\beta$ this observable is 
continuous across $\mu^2=0$, which is a good indication that complex 
Langevin dynamics works well in this region. The cusp at 
$\mu_\rmI=\pi/N_\tau$ (corresponding to 
$\mu^2=-0.154$) reflects the Roberge-Weiss transition.
 At lower $\beta$, however, we observe that the action density is no 
longer continuous: this is interpreted as a breakdown of complex Langevin 
dynamics.
 In order to verify this, Fig.\ \ref{fig:action-plots} also contains the 
expectation values of the action density found using the worm algorithm in 
the world line formalism for real $\mu$. As expected, in this case the 
action density is continuous across $\mu^2=0$ for all values of $\beta$, 
confirming the interpretation given above.
 We have verified that the jump in the action density at lower $\beta$ is 
independent of the lattice volume. We have also verified that the 
discrepancy at $\mu^2=0$ between real Langevin dynamics and the world line 
result (at e.g.\ $\beta=0.4$) is due to the finite Langevin stepsize.

For small $\beta$, the numerical results found with the worm algorithm are 
consistent with those derived analytically in the strong-coupling limit 
above. The expectation value of the action density is $\mu$ independent 
and hence the Roberge-Weiss periodicity is smoothly realized. Using Eq.\ 
(\ref{eqSsc}), we also find quantitative agreement: in the strong-coupling 
expansion $\bra S\ket/\Omega = -0.0621 +{\cal O}(10^{-4})$ for $\beta=0.2$ 
and $-0.145 +{\cal O}(10^{-3})$ for $\beta=0.3$.

As discussed above, for the parameter values and lattice sizes used here 
the sign problem is not severe: taking 
$\mu^2=0.1$ and $\beta=0.2$, we find that 
 \be 
 \Omega\Delta f \approx \Omega \frac{\beta^2\mu^2}{4} \approx 0.51,
\;\;\;\;\;\;\;\;\;\;\;\;
 \bra e^{i\varphi}\ket_{\rm pq} \approx 0.60. 
 \ee 
 We take this as a first indication that the observed breakdown is not due 
to the presence of the sign problem, especially since complex Langevin 
dynamics has been demonstrated to work well in other models where the sign 
problem is severe \cite{Aarts:2008wh,Aarts:2009hn}.

\begin{figure}[t]
\centering
\includegraphics[width=0.85\textwidth]{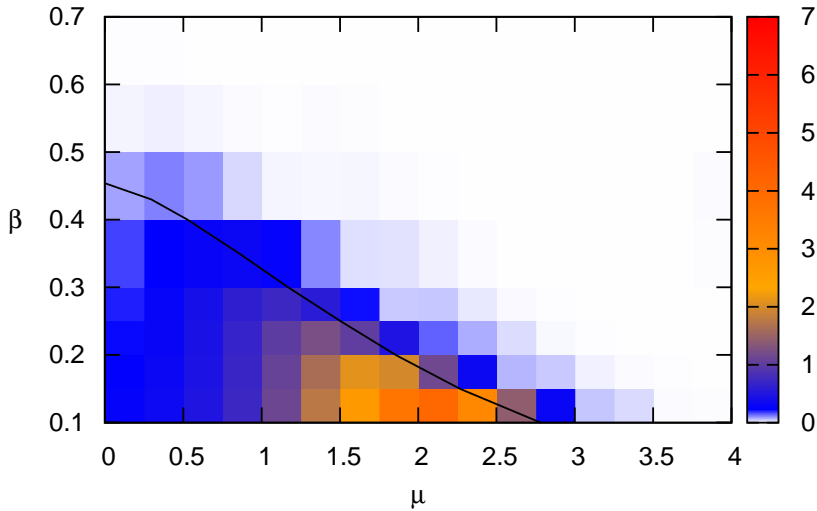}
 \caption{Colour plot indicating the relative difference $\Delta S$ 
between the expectation value of the action density obtained with complex 
Langevin dynamics and in the world line formulation, see Eq.\ 
(\ref{eq:relS}). Also shown is the phase boundary $\beta_c(\mu)$ between 
the ordered (large $\beta$) and disordered (small $\beta$) phase 
\cite{Banerjee:2010kc}.
 }
\label{fig:percent}
\end{figure}
  
To probe the reliability of complex Langevin dynamics for larger 
values of $\mu$, we have computed the action density for a large number 
of parameter values in the $\beta-\mu$ plane. Our findings are summarized 
in Fig.\ \ref{fig:percent}, where we show the relative difference between 
the action densities obtained with complex Langevin (cl) and in the world 
line formulation (wl), according to
 \be
\label{eq:relS}
\Delta S = \frac{\bra S\ket_{\rm wl} - \bra S\ket_{\rm cl}}{\bra 
S\ket_{\rm wl}}.
\ee
Also shown in this figure is the phase transition line $\beta_c(\mu)$, 
taken from Ref.\ \cite{Banerjee:2010kc}. We observe a clear correlation 
between the breakdown of Langevin dynamics and the phase boundary: complex 
Langevin  dynamics works fine well inside the ordered phase, but breaks 
down in the boundary region and the disordered phase. 
The largest deviation around $\mu=2$ is due to the Silver Blaze effect: 
the difference between the action density found with complex Langevin 
dynamics and the correct $\mu$-independent action density is maximal just 
before crossing over to the other phase, where the agreement improves 
quickly.


\section{Diagnostics}
\label{sec:diagnostics}
\setcounter{equation}{0}

In this section we attempt to characterize the results presented above in 
terms of properties of complex Langevin dynamics and the distribution 
$P[\phi^\rmR, \phi^\rmI]$ in the complexified field space, see Eq.\ 
(\ref{eqP}). We suppress Langevin time dependence, since we always 
consider the quasi-stationary regime, i.e.\ the initial part of the 
evolution is discarded (we considered Langevin times up to $\vartheta \sim 
2\times 10^4$). Our aim is to argue that the discrepancy at small $\beta$ 
is introduced by complex Langevin dynamics rather than by the presence of 
a chemical potential and hence not due to the sign problem.

A first test of the validity of complex Langevin dynamics is to compare 
simulations at $\mu=0$ using a cold start, i.e.\ with $\phi^\rmI=0$ 
initially, and a hot start in which $\phi^\rmI$ is taken from a Gaussian 
distribution.\footnote{The real components $\phi^\rmR$ are taken from a 
Gaussian distribution always.} When $\mu=0$, a cold start corresponds to 
real Langevin dynamics. In the case of a hot 
start, however, the fields lie immediately in the complexified space and 
so the dynamics is complexified. Comparison of results obtained with these 
two initial ensembles gives insight into the inner workings of 
complex Langevin dynamics.
 We have computed the expectation value of the action density at $\mu=0$
using both a hot and a cold start. We found them to agree at large
$\beta$, despite the fact that the imaginary components of the field are
initialised randomly. However, when $\beta\lesssim 0.5$, they disagree. 
Moreover, the result from the cold start agrees with the one obtained in
the world line formulation. One is therefore led to conclude that when
$\mu=0$ the imaginary components $\phi^\rmI$ are driven to zero (more 
precisely, to a constant value)
at large
$\beta$ but are not constrained at small $\beta$. In other words the
drift terms are not capable of restoring the reality of the dynamics. It
is tempting to relate this to being in (or close to) the disordered
phase. We note that it cannot be understood from the classical fixed
point structure, since this is independent of $\beta$. We also remark
that the dynamics at small $\beta$ resembles Langevin dynamics with
complex noise ($N_\rmI>0$) \cite{Aarts:2009uq}, where the trajectories are
kept in the complexified field space by the stochastic kicks on
$\phi^\rmI$ (rather than by the drift terms, as is the case here).

\begin{figure}[t]
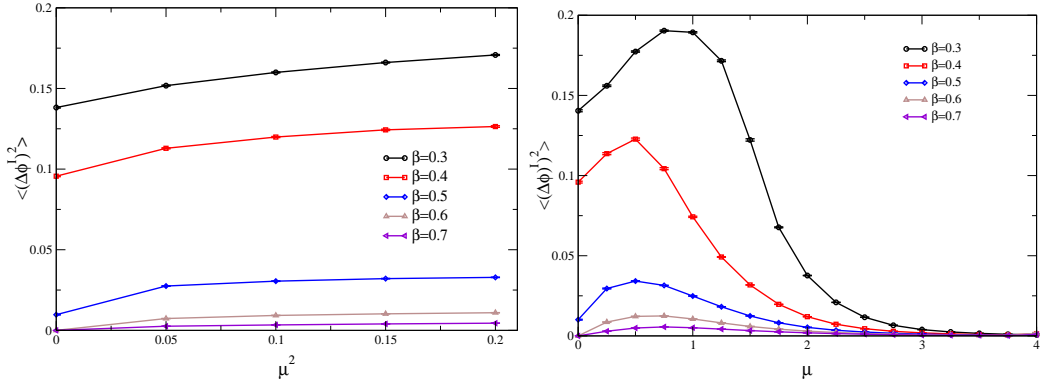

\centering
\includegraphics[width=0.47\textwidth]{img2-i2_x10.eps}
\includegraphics[width=0.47\textwidth]{img2-i2_x8.eps}
 \caption{Width of the distribution $P[\phi^\rmR,\phi^\rmI]$ in the 
imaginary direction for various values of $\beta$ 
as a function of $\mu^2$ on a $10^3$ lattice (left) and,  
for larger $\mu$, as a function of $\mu$ on a $8^3$ lattice (right). 
 }
\label{fig:imag2}
\end{figure}

In terms of the distribution $P[\phi^\rmR, \phi^\rmI]$, these findings 
imply that $P[\phi^\rmR, \phi^\rmI]\sim e^{-S}\delta(\phi^\rmI)$ at large 
$\beta$, but not at small $\beta$. This can be further investigated by 
studying the width of the distribution in the imaginary 
direction,\footnote{The mean value $\bra\phi^\rmI\ket=0$; in the large 
$\beta$ phase, this requires averaging over a large number of initial 
conditions.} 
\be
 \left\langle \left(\Delta\phi^{\rmI}\right)^2 \right\rangle = 
  \left\langle \frac{1}{\Omega}\sum_x \left(\phi_{x}^\rmI\right)^2 
 \right\rangle
  - \left\langle \frac{1}{\Omega}\sum_x \phi_{x}^\rmI \right\rangle^2. 
\ee 
 When $\mu=0$ the width should vanish, while when turning on $\mu$ one may 
expect it to increase smoothly. The results are shown in 
Fig.~\ref{fig:imag2}. For the larger $\beta$ values this is exactly what 
is observed: the width increases smoothly from zero. For the smaller 
$\beta$ values, however, we observe that the width is nonzero even when 
$\mu=0$ (when using a hot start), and remains large for nonzero $\mu$. At 
larger values of $\mu$ the width is driven again towards zero and 
agreement with the world line results improves, see Fig.\ 
\ref{fig:percent}. We remark here that it is possible that different 
distributions (with different widths) yield the same result for 
observables. This is what is theoretically expected in the presence of 
complex noise ($N_\rmI\geq 0$) \cite{Aarts:2009uq} and can be seen 
analytically in gaussian models with complex noise, where a continuous 
family of distributions $P[\phi^\rmR,\phi^\rmI;N_\rmI]$ all yield the same 
result for observables, independent of $N_\rmI$, even though the width of 
these distributions is nonzero and increases with $N_\rmI$ \cite{gert}. In 
the case we study here, however, we find that the failure of complex 
Langevin dynamics in the disordered phase is correlated with the spread of 
the distribution $P[\phi^\rmR,\phi^\rmI]$ in the noncompact direction. We 
conclude that a relatively narrow distribution, with a smoothly increasing 
width, is required. We note again that this resembles observations made in 
simulations of nongaussian models with complex noise 
\cite{Aarts:2009uq,seiler}.

To investigate the interplay between (the width of) the distribution and 
observables, we express expectation values as
 \be
\bra A[\phi^\rmR, \phi^\rmI] \ket = \frac{1}{Z}\int D\phi^\rmR 
D\phi^\rmI\, P[\phi^\rmR, \phi^\rmI]A[\phi^\rmR, \phi^\rmI],
\ee
with
\be
 Z = \int D\phi^\rmR D\phi^\rmI\, P[\phi^\rmR, \phi^\rmI].
\ee
 In general the operator $A$ is not required to be holomorphic, i.e.\ a 
function of $\phi^\rmR+i\phi^\rmI$, since this will allow more 
insight in properties of the distribution.\footnote{Of course 
only holomorphic functions correspond to observables in the original 
theory.}
 The distribution of an operator $A$ can then be defined according to
\be
 \bra A \ket = \int dA\, P(A)A = \frac{1}{Z}\int D\phi^\rmR D\phi^\rmI\, 
 P[\phi^\rmR, \phi^\rmI]A[\phi^\rmR, \phi^\rmI],
\ee
 where 
\be
 P(A) = \frac{1}{Z}\int D\phi^\rmR D\phi^\rmI\, P[\phi^\rmR, \phi^\rmI]
\delta(A - A[\phi^\rmR, \phi^\rmI]),
\ee
with the normalization
 \be
\int dA \, P(A) = 1.
\ee 
Distributions $P(A)$ can be constructed numerically, by sampling $A$ from 
configurations generated by complex Langevin dynamics.

\begin{figure}[t]
\centering
\includegraphics[width=0.7\textwidth]{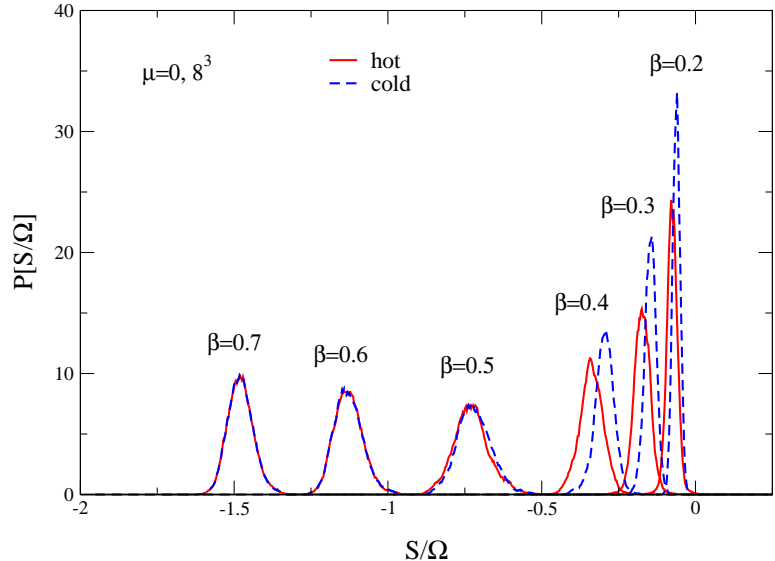}
\caption{Distribution of action density $S/\Omega$ for various values of 
$\beta$ at $\mu=0$ on a $8^3$ lattice, using a hot and a cold start.}
\label{fig:act-dens}
\end{figure}

The distribution for the action density is shown in 
Fig.~\ref{fig:act-dens}, comparing again a hot and cold start at $\mu=0$. 
This figure supports the earlier claim that real and complex Langevin 
match at larger $\beta$ but fail at smaller $\beta$. However, the reason 
for failure is somewhat subtle. Na\"{\i}vely, one might expect a large 
``tail'' caused by excursions in the complexified field space to affect the 
expectation value but this does not appear to happen. Instead we find that 
the entire distribution is shifted and becomes only slightly wider at 
$\beta\lesssim 0.5$ when the hot start is used.

\begin{figure}[!p]
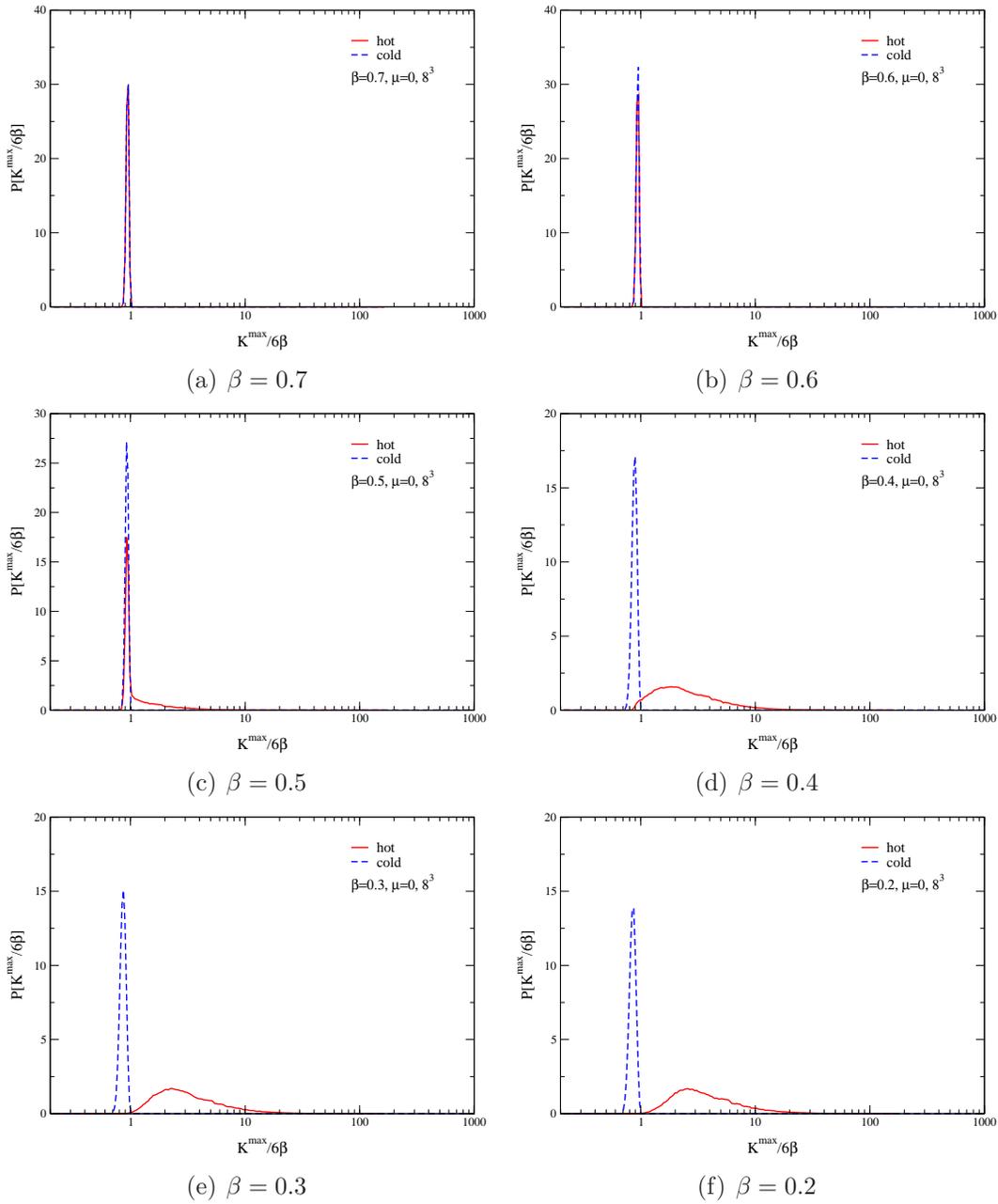

\centering
\subfigure[$\beta=0.7$]{
\includegraphics[width=0.47\textwidth]{prob_kmax_b0.7_x8.eps}
}
\subfigure[$\beta=0.6$]{
\includegraphics[width=0.47\textwidth]{prob_kmax_b0.6_x8.eps}
}
\subfigure[$\beta=0.5$]{
\includegraphics[width=0.47\textwidth]{prob_kmax_b0.5_x8.eps}
}
\subfigure[$\beta=0.4$]{
\includegraphics[width=0.47\textwidth]{prob_kmax_b0.4_x8.eps}
}
\subfigure[$\beta=0.3$]{
\includegraphics[width=0.47\textwidth]{prob_kmax_b0.3_x8.eps}
}
\subfigure[$\beta=0.2$]{
\includegraphics[width=0.47\textwidth]{prob_kmax_b0.2_x8.eps}
}
 \caption{Distribution of $K^{\rm max}/(6\beta)$ at $\mu=0$ on a $8^3$ 
lattice 
using a hot and a cold start.
}
\label{fig:prob-kmax}
\end{figure}
\begin{figure}[!p]
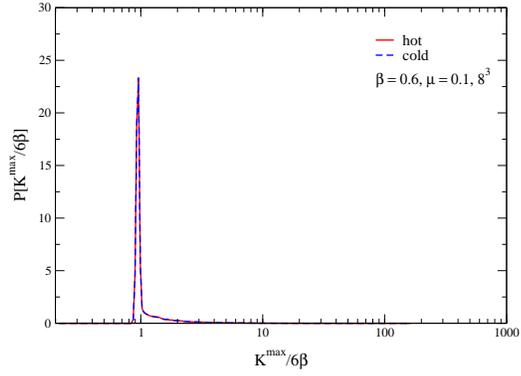
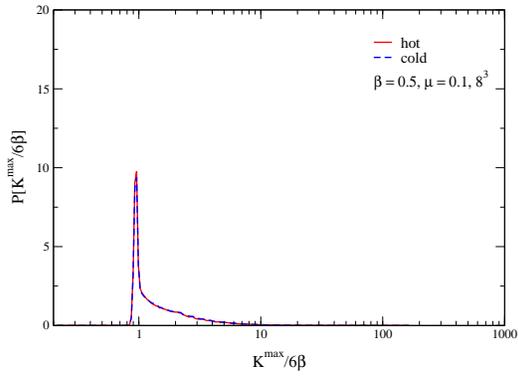
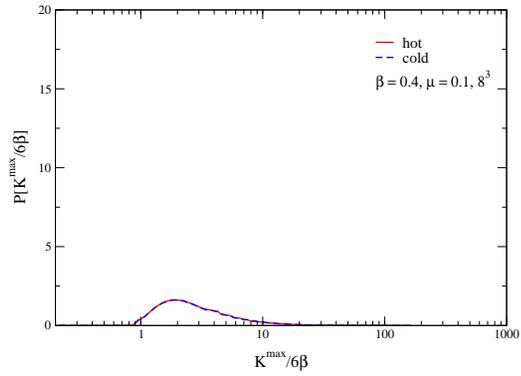
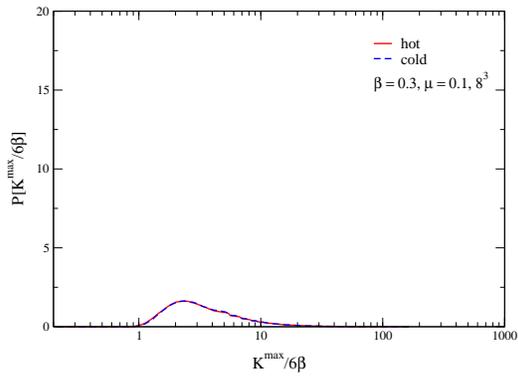
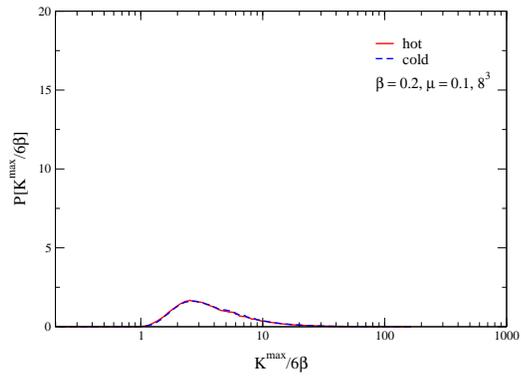

\centering
\subfigure[$\beta=0.7$]{
\includegraphics[width=0.47\textwidth]{prob_kmax_b0.7_mu0.1_x8.eps}
}
\subfigure[$\beta=0.6$]{
\includegraphics[width=0.47\textwidth]{prob_kmax_b0.6_mu0.1_x8.eps}
}
\subfigure[$\beta=0.5$]{
\includegraphics[width=0.47\textwidth]{prob_kmax_b0.5_mu0.1_x8.eps}
}
\subfigure[$\beta=0.4$]{
\includegraphics[width=0.47\textwidth]{prob_kmax_b0.4_mu0.1_x8.eps}
}
\subfigure[$\beta=0.3$]{
\includegraphics[width=0.47\textwidth]{prob_kmax_b0.3_mu0.1_x8.eps}
}
\subfigure[$\beta=0.2$]{
\includegraphics[width=0.47\textwidth]{prob_kmax_b0.2_mu0.1_x8.eps}
}
 \caption{As in the previous figure, for $\mu=0.1$}
\label{fig:prob-kmax2}
\end{figure}

Finally, the observed difference at large and small $\beta$ also appears 
prominently in the actual dynamics, i.e.\ in the drift 
terms. We have analyzed the maximal force $K^{\rm max}$ appearing in the 
adaptive stepsize algorithm. In the case of real Langevin dynamics, the 
drift terms are limited by an upper bound of $K^{\rm max} \le 6\beta$. In 
the complexified space there is no upper limit and the drift terms can in 
principle become several orders of magnitude larger~\cite{Aarts:2009dg}. 
The distribution of $K^{\rm max}$ is plotted at $\mu=0$ with hot and cold 
starts in Figure~\ref{fig:prob-kmax}. In the large $\beta$ phase, the 
distributions appear identical, with $K^{\rm max}\le 6\beta$. This is 
consistent with the conclusion reached above. In the low $\beta$ phase the 
distributions are dramatically different: in the complexified dynamics, 
triggered by the hot start, much larger forces appear. The distributions 
are no longer peaked but very broad with a long tail (note the horizontal 
logarithmic scale). At $\beta=0.5$ we observe interesting crossover 
behaviour: both the peaked distribution bounded by $K^{\rm max} = 6\beta$ 
and a decaying ``tail'' characteristic of small $\beta$ distributions 
appear.

To study the two possible distributions of $K^{\rm max}$ further, we show 
in Fig.~\ref{fig:prob-kmax2} the same results but now with $\mu=0.1$. In 
this case the hot and cold start yield identical distributions, since both 
simulations are complexified due to the nonzero chemical potential. The 
striking difference between the distributions at large and small $\beta$ 
is still present. At large $\beta$ the force can occasionally be large, 
making the use of an adaptive stepsize necessary. However, the typical 
value is still determined by the maximal value for real Langevin dynamics, 
i.e.\ $6\beta$. At small $\beta$ this part of the distribution is 
completely gone and is replaced by a broad distribution at much larger 
$K^{\rm max}$ values. Again at $\beta=0.5$ we observe crossover behaviour 
with both features present. These results are qualitatively the same on 
larger volumes.

Let us summarize the findings of this section. Complex Langevin dynamics
works well at large $\beta$ in the ordered phase. The distribution
$P[\phi^\rmI,\phi^\rmI]$ in the complexified field space is relatively
narrow in the noncompact direction and Langevin simulations started with
hot and cold initial conditions agree. The drift terms do occasionally
become large but the typical size is set by the maximal value for real
Langevin evolution.
 At small $\beta$, in or close to the disordered phase, the distribution
is much wider in the $\phi^\rmI$ direction. Typical drift terms are
much larger, with a wide spread in the distribution. At $\mu=0$
complexified dynamics does not reduce to real dynamics. There is a
strong correlation with the phase the theory is in (see Fig.\
\ref{fig:percent}), but not with the sign problem, since these
observations also hold at $\mu=0$ and are independent of the lattice 
volume. Moreover, for the lattice volumes we consider the sign problem is 
not severe. 
 We emphasize that a firm conclusion can only be drawn after all the
findings presented above are combined consistently, while the observation
of e.g.\ large drift terms or a large width by itself would clearly be
insufficient.

\section{Conclusion}
\label{sec:conclusion}
\setcounter{equation}{0}

We have studied the applicability of complex Langevin dynamics to simulate 
field theories with a complex action due to a finite chemical potential, 
in the case of the three-dimensional XY model. Using analytical 
continuation from imaginary chemical potential and comparison with the 
world line formulation we found that complex Langevin dynamics yields 
reliable results at larger $\beta$ but fails when $\beta\lesssim 0.5$ at 
small chemical potential. We established that the region of failure is 
strongly correlated with the part of the phase diagram which corresponds 
to the disordered phase. We have verified that these conclusions do not 
depend on the lattice volume. Failure at small $\beta$ values was also 
observed a long time ago in the case of SU(3) field theory in the presence 
of static charges \cite{Ambjorn:1986fz}.

Due to the use of an adaptive stepsize algorithm no runaways or 
instabilities have been observed. The results we found in the disordered 
phase are therefore interpreted as convergence to the wrong result. To 
analyze this, we have studied properties of the dynamics and field 
distributions in the complexified field space. For the smaller $\beta$ 
values, we found that complexified dynamics does not reduce to real 
dynamics when $\mu=0$. Furthermore, for the system sizes and parameter 
values we used, the sign problem is not severe. We conclude therefore that 
the failure is not due to the presence of the sign problem, but rather due 
to an incorrect exploration of the complexified field space by the 
Langevin evolution. The forces appearing in the stochastic process behave 
very differently at large and small $\beta$.  Interestingly, in the 
crossover region at $\beta\approx 0.5$, the dynamics shows a combination 
of large and small $\beta$ characteristics. It would be interesting to 
further understand this, e.g.\ in terms of competing (nonclassical) fixed 
points.

We found that several features resemble those found in simulations of 
simple models with complex noise \cite{Aarts:2009uq,seiler}. Our hope is 
therefore that a detailed study of simple models with complex noise can 
shed light on the features observed here with real noise. Such an 
investigation is currently in progress.


 \vspace*{0.5cm}
 \noindent
 {\bf Acknowledgments.}

\noindent
 We thank Chris Allton and Simon Hands, Debasish Banerjee and Shailesh 
Chandrasekharan, Philippe de Forcrand, and especially Kim Splittorff, 
Erhard Seiler and Ion-Olimpiu Stamatescu for discussions. We thank the 
Blue C Facility at Swansea University for computational resources. This 
work is supported by STFC.

\end{document}